\documentstyle[12pt]{article}
\newcommand{\be}{\begin{equation}}
\newcommand{\ee}{\end{equation}}
\newcommand{\bea}{\begin{eqnarray}}
\newcommand{\eea}{\end{eqnarray}}
\newcommand{\ba}{\begin{array}}
\newcommand{\ea}{\end{array}}

\newcommand{\norsl}{\normalsize\sl}
\newcommand{\norsc}{\normalsize\sc}
\begin{document}

\begin{titlepage}

\title{
Target Mass Effects in QCD Bjorken Sum Rule
\\
\quad \\
\quad \\
}

\author{
\norsc  Hiroyuki KAWAMURA and
         Tsuneo UEMATSU\thanks{Supported in part by
          the Monbusho Grant-in-Aid for Scientific Research
          No. C-06640392 and \quad No.06221239.} \\
\quad \\
\norsl  Department of Fundamental Sciences\\
\norsl  FIHS, Kyoto University\\
\norsl  Kyoto 606-01, JAPAN}

\date{}
\maketitle

\vspace{4cm}

\begin{abstract}
{\normalsize
We investigate the target mass effects in QCD Bjorken sum rule. The magnitude
of the target mass correction is estimated in a variety of methods employing
positivity bound as well as the experimental data for the asymmetry
parameters.
It turns out that the target mass correction is sizable
at low $Q^2$ of the order of a few GeV$^2$, where the QCD correction is
significant. We show that there exists uncertainty due to target mass effects
in determining the QCD effective coupling constant $\alpha_s(Q^2)$ from the
Bjorken sum rule.
}
\end{abstract}

\begin{picture}(5,2)(-335,-670)
\put(2.3,-125){KUCP-74}
\put(2.3,-140){September 1994}
\end{picture}

\vspace{2cm}

\thispagestyle{empty}
\end{titlepage}
\setcounter{page}{1}
\baselineskip 24pt


There has been a lot of interest in the nucleon
spin structure functions which can be measured by the deep inelastic
scattering of polarized leptons on polarized nucleon targets \cite{SLAC,EMC}.
The nucleon spin structure is described by the two spin
structure functions $g_1(x,Q^2)$ and $g_2(x,Q^2)$.
Recent experiments on the $g_1(x,Q^2)$ for the deuteron, $^3$He and proton
targets at CERN and SLAC \cite{SMCD,E142,SMCP} together with
EMC data \cite{EMC}
have provided us with the data for testing the Bjorken sum rule \cite{Bj} as
well as the Ellis-Jaffe sum rule \cite{EJ}, and also for studying $Q^2$
evolution of $g_1(x,Q^2)$ \cite{ANR}.

In the framework of the operator product expansion and the renormalization
group, the Bjorken sum rule with QCD radiative corrections reads:
\be
\int_0^1 dx \bigl [ g_1^p(x,Q^2)-g_1^n(x,Q^2) \bigr ]
=\frac{1}{6} \frac{G_A}{G_V} \bigl [ 1-\frac{\alpha_s(Q^2)}{\pi}+
{\cal O}(\alpha_s^2) \bigr ].
\label{bj}
\ee
where $g_1^p(x,Q^2)$ and $g_1^n(x,Q^2)$ are the spin structure function of
proton and neutron, respectively, with $x$ and $Q^2$ being the Bjorken
variable and the virtual photon mass squared. On the right-hand side,
$G_A/G_V \equiv g_A$ is the ratio of the axial-vector to vector coupling
constants.
The first order QCD correction was calculuated in \cite{KMMSU,KMSU,K} and
the higher order corrections were given in \cite{GL,LV,LTV,KS}.

In order to confront the QCD prediction with the experimental data at low
$Q^2$ where the QCD corrections are significant, we have to take into account
the corrections due to the mass of the target.  In this $Q^2$ region where we
cannot neglect the order $M^2/Q^2$ terms, with $M$ being the nucleon mass, we
have to extract the definite spin contribution in the operator product
expansion. This can be achieved by considering the Nachtmann moments of the
structure functions \cite{N}.

Some years ago, the target mass effects for polarized deep inelastic
scattering were studied in refs. \cite{WANDZ,MU}.
The Nachtmann moments for the twist two and three operators in
operator product expansion relevant to polarized deep inelastic scattering
were obtained in closed analytic forms in ref. \cite{MU}.
Taking the first Nachtmann moment we can extract the contribution from the
spin 1 and twist-2 operator, and
the Bjorken sum rule with target mass corrections reads \cite{MU}:
\bea
& &{1\over 9}\int_0^1 dx {{\xi^2}\over{x^2}} \Bigl [
5+ 4\sqrt{1+{{4M^2x^2}\over{Q^2}}} \Bigr ]
\Bigl [ g_1^p(x,Q^2)-g_1^n(x,Q^2)\Bigr ] \nonumber\\
& &-{4\over 3}\int_0^1 dx {{\xi^2}\over{x^2}}{{M^2x^2}\over{Q^2}}
\Bigl [ g_2^p(x,Q^2)-g_2^n(x,Q^2)\Bigr ]
={1\over 6}{G_A\over G_V}\Bigl [1-{{\alpha_s(Q^2)}\over \pi}+O(\alpha_s^2)
\Bigr ] ,
\label{nacht}
\eea
where $M$ denotes the nucleon mass and the variable $\xi$ \cite{GP} is given by
\be
\xi= {{2x}\over{1+\sqrt{1+4M^2x^2/Q^2}}}.
\label{xi}
\ee
Note that in the presence of target mass correction, the other spin structure
function $g_2^{p,n}(x,Q^2)$ also comes into play in the Bjorken sum rule.
It should also be noted that target mass corrections considered as the
expansion in powers of $M^2/Q^2$ is not valid when $M^2/Q^2$ is of order
unity \cite{SV,BBK}.

Taking the difference between the left-hand side of (\ref{nacht}) and that of
(\ref{bj}), we get the target mass correction $\Delta\Gamma$ due to the
necessary projection onto the definite spin as
\bea
\Delta\Gamma &=&\int_0^1dx \bigl \{\frac{5}{9}\frac{\xi^2}{x^2}+
\frac{4}{9}\frac{\xi^2}{x^2}\sqrt{1+\frac{4M^2x^2}{Q^2}}-1\bigr\}
\times \Bigl [ g_1^p(x,Q^2) - g_1^n(x,Q^2) \Bigr ] \nonumber\\
& &-{4\over 3}\int_0^1 dx {{\xi^2}\over{x^2}}{{M^2x^2}\over{Q^2}}
\Bigl [ g_2^p(x,Q^2)-g_2^n(x,Q^2)\Bigr ].
\eea

We now study the size of the target mass correction $\Delta\Gamma$ to
the Bjorken sum rule.
First we note that in terms of virtual photon asymmetry parameters $A_1$ and
$A_2$, the structure functions are given by
\bea
g_1(x,Q^2)&=&{{F_2(x,Q^2)[A_1(x,Q^2)+ \gamma A_2(x,Q^2)]}\over
{2x[1+R(x,Q^2)]}}, \nonumber\\
g_2(x,Q^2)&=&{{F_2(x,Q^2)[-A_1(x,Q^2)+ A_2(x,Q^2)/\gamma]}\over
{2x[1+R(x,Q^2)]}},
\label{g1g2}
\eea
where $F_2(x,Q^2)$ is the unpolarized structure function, $\gamma=
\sqrt{4M^2x^2/Q^2}$ and
$R(x,Q^2)=\sigma_L(x,Q^2)/\sigma_T(x,Q^2)$ is the ratio of
the longitudinal to transverse virtual photon cross sections.

The target mass correction turns out to be
\bea
\Delta\Gamma&=&\int_0^1 dx \Bigl \{{5\over 9}{\xi^2\over x^2}
+{4\over 9}{\xi^2\over x^2}\sqrt{1+4M^2x^2/Q^2}-1 \Bigr \} \nonumber\\
& &\qquad \times {1\over{2x}}\Bigl \{
{{F_2^p(A_1^p+\gamma A_2^p)}\over{1+R^p}}
-{{F_2^n(A_1^n+\gamma A_2^n)}\over{1+R^n}}\Bigr \} \nonumber\\
& &-{4\over 3}\int_0^1 dx{\xi^2\over x^2}{{M^2x^2}\over{Q^2}}
{1\over{2x}}\Bigl \{
{{F_2^p(-A_1^p+A_2^p/\gamma)}\over{1+R^p}}
-{{F_2^n(-A_1^n+A_2^n/\gamma)}\over{1+R^n}}\Bigr \}.
\label{delta}
\eea

We estimate the target mass correction $\Delta\Gamma$ in a variety of methods.
First of all, we apply the positivity bound for the asymmetry parameters
\cite{DDR}:
\be
|A_1| \leq 1, \qquad |A_2| \leq \sqrt{R},
\label{a1a2}
\ee
to get the upper bound for $\Delta\Gamma$ (Analysis I).

In this case we have
\bea
|\Delta\Gamma| &\leq&
\int_0^1 dx{1\over{2x}} \Bigl |{5\over 9}{\xi^2\over x^2}
+{4\over 9}{\xi^2\over x^2}\sqrt{1+{4M^2x^2\over Q^2}}
+{4\over 3}{\xi^2\over x^2}{{M^2x^2}\over{Q^2}}-1 \Bigr |
\times ({{F_2^p}\over{1+R^p}}+{{F_2^n}\over{1+R^n}}) \nonumber\\
&+&\int_0^1 dx{1\over{2x}} \Bigl |({5\over 9}{\xi^2\over x^2}
+{4\over 9}{\xi^2\over x^2}\sqrt{1+{4M^2x^2\over Q^2}}-1)
\sqrt{4M^2x^2/Q^2} \nonumber\\
& &-{4\over 3}{\xi^2\over x^2}{{M^2x^2}\over{Q^2}}
{1\over\sqrt{4M^2x^2/Q^2}} \Bigr | \times
({{F_2^p\sqrt{R^p}}\over{1+R^p}}+{{F_2^n\sqrt{R^n}}\over{1+R^n}}).
\label{deltagam}
\eea

Using the parametrization for $R$ taken from the global fit of the SLAC data
\cite{RSLAC} and the NMC parametrization for $F_2(x,Q^2)$ \cite{NMC},
we obtain for the average $Q^2$ of the E142 data, $Q^2=$ 2.0 GeV$^2$ and the
SMC data, $Q^2=$ 4.6 GeV$^2$;
\be
|\Delta\Gamma|\leq 0.036 \quad \mbox{for $Q^2=2.0$ GeV$^2$}, \qquad
|\Delta\Gamma|\leq 0.0177 \quad \mbox{for $Q^2=4.6$ GeV$^2$},
\label{SMC}
\ee
whereas the value of the SMC experiment is $\Gamma\equiv \Gamma_1^p-\Gamma_1^n
=0.20\pm 0.05\pm 0.04(\mbox{\rm syst.})$, where $\Gamma_1^{p(n)}=
\displaystyle{\int_0^1dx g_1^{p(n)}(x,Q^2)}$ \cite{EK1}.
In (\ref{SMC}), the error of the upper bounds of $\Delta\Gamma$
due to the parametrizations $R$ and $F_2$ is expected to be around 10 $\%$.

In Fig.1 we have plotted the upper bound for $\Delta\Gamma$,
which we denote by $\Delta\Gamma_{u.b.}$ (i.e. $|\Delta\Gamma| \leq
\Delta\Gamma_{u.b.}$), as a function of
$Q^2$ for Analysis I by a solid line.

So far we have not made use of any experimental data on spin asymmetries
$A_1$ and $A_2$. Now we employ the experimetal data to improve the upper bound.
We take the data on $A_1^p$ from SMC data \cite{SMCP} together with EMC
data \cite{EMC} and those for $A_1^d$ from SMC group \cite{SMCD} to extract
$A_1^n$, for which we can also use the E142 data \cite{E142}. As for the
$A_2^{p,n}$ there are two possibilities : i) Either use the positivity bound
$|A_2| \leq \sqrt{R}$ both for the proton and the neutron, or ii) use the
recently measured $A_2^p$ by the SMC group \cite{SMCA2} together with the
positivity bound for $A_2^n$.

For the choice i) which we call Analysis II, the upper bound for the
$\Delta\Gamma$, $\Delta\Gamma_{u.b.}$, is shown in Fig.1 by the short-dashed
line, which is located slightly lower than $\Delta\Gamma_{u.b.}$ for
Analysis I. This situation can be understood by the following observation.
If we decompose the $\Delta\Gamma_{u.b.}$ into two parts, $\Delta\Gamma_1$
and $\Delta\Gamma_2$, which are the contributions from $A_1$ and $A_2$,
respectively, it turns out that $\Delta\Gamma_2$ is much larger than
$\Delta\Gamma_1$. The value of $\Delta\Gamma_1$ turns out be less than 10
$\%$ of $\Delta\Gamma_2$. Furthermore we note that the integrals are not so
sensitive to the parametrization of $A_1$, if we assume the Regge behavior
for $x \sim 0$, and $A_1 \rightarrow 1$ for $x \rightarrow 1$.
Here we have also assumed that there is no significant $Q^2$ dependence in
$A_1$ as observed in the experiments [1-5].

For the choice ii) which will be called Analysis III, we have also plotted
the upper bound in Fig.1 by the long-dashed line. Here we took the data
on $A_2^p$ obtained by SMC group at the first measurement of transverse
asymmetries \cite{SMCA2}, where the number of data points are still four and
the relative error bars are not so small. The $A_2^p$ measured is much
smaller than the positivity bound.
If the $A_2$ for the neutron is also small as mentioned in ref.\cite{E142},
the $\Delta\Gamma_{u.b.}$ becomes very small.

Now we turn to the issue related to the determination of the QCD coupling
constant from Bjorken sum rule which has recently been discussed by Ellis
and Karliner \cite{EK2}.
Up to the ${\cal O}(\alpha_s^4)$ we have the following QCD corrections
\cite{LV,LTV,KS}:
\vspace{0.5cm}
\bea
\Gamma(Q^2)&=&\frac{1}{6}\frac{G_A}{G_V} \Bigl [
1-\frac{\alpha_s(Q^2)}{\pi}-3.5833\bigl (\frac{\alpha_s(Q^2)}{\pi}
\bigr )^2 \nonumber\\
& &-20.2153\bigl (\frac{\alpha_s(Q^2)}{\pi}\bigr )^3 -{\cal O}(130)
\bigl (\frac{\alpha_s(Q^2)}{\pi}\bigr )^4 +\cdots \Bigr ].
\label{gamqcd}
\eea
By putting $\frac{1}{6}{G_A}/{G_V}=0.2095$ and taking the left-hand side
of (\ref{gamqcd}) at $Q^2=2.5$GeV$^2$ to be 0.161, which was obtained
by Ellis and Karliner
in their analysis of E142 and E143 data \cite{EK2}, we find
$\alpha_s(Q^2=2.5\mbox{GeV}^2)=0.375$ \cite{EK2}. The $Q^2=2.5\mbox{GeV}^2$
is the averaged value of the mean $Q^2$ of the E142 data
($<Q^2>\simeq 2\mbox{GeV}^2$) and the preliminary E143 data
($<Q^2>\simeq 3\mbox{GeV}^2$).

Here we shall not take into account the higher-twist effects which are
considered to be rather small as claimed in refs. \cite{EK2}.

In Fig.2 we have shown the QCD coupling constant $\alpha_s$ as a function of
$\Gamma$. Here we note that $\alpha_s$ varies significantly with the change
of the $\Gamma$.
The uncertainty in $\Gamma$ due to target mass effects gives rise to that
for the QCD coupling constant $\alpha_s(Q^2=2.5\mbox{GeV}^2)$.
Namely, for the variation
\be
0.132 \leq \Gamma \leq 0.190 \quad (\mbox{Analysis I}),
\ee
we obtain the ambiguity for $\alpha_s$
\be
0.213 \leq \alpha_s(Q^2=2.5\mbox{GeV}^2) \leq 0.474 \quad (\mbox{Analysis I}).
\ee
For the analyses II and III, we have
\be
0.134 \leq \Gamma \leq 0.188 \quad (\mbox{Analysis II}), \qquad
0.148 \leq \Gamma \leq 0.174 \quad (\mbox{Analysis III}),
\ee
which lead to the ambiguity for $\alpha_s$
\bea
0.228 &\leq& \alpha_s(Q^2=2.5\mbox{GeV}^2) \leq 0.469 \quad
(\mbox{Analysis II}),
\nonumber\\
0.315 &\leq& \alpha_s(Q^2=2.5\mbox{GeV}^2) \leq 0.424 \quad
(\mbox{Analysis III}).
\eea


To summarize, in this paper we have examined the possible corrections to the
Bjorken sum rule coming from target mass effects.
We have found that at relatively small $Q^2$ where the QCD
effect is significant, the target mass effects are also non-negligible.
We found that to test the target mass correction precisely, we need
accurate data for $A_2(x,Q^2)$.
In determining the QCD coupling constant $\alpha_s$ from the Bjorken sum
rule, there appears uncertainty due to target mass effects.
This uncertainty can also be removed by the experimental data on $A_2(x,Q^2)$.

Although in this paper we have confined ourselves to the target mass effects
in the Bjorken sum rule, the similar analysis can be carried out for the
Ellis-Jaffe sum rule which will be discussed elsewhere \cite{KU}.

We hope that future experiments at CERN, SLAC and DESY will provide us with
data on $A_1$ possessing higher statistics as well as the data on $A_2$
with high accuracy which will enable us to study $g_2$ structure functions
and also target mass effects more in detail.

\vspace{2 cm}

 The authors would like to thank J. Kodaira, S. Matsuda and Y. Mizuno
for valuable discussions.
Part of this work was done while one of us (T.U.) was staying at DESY
in the summer of 1993. He thanks the DESY Theory Group for its hospitality.

\vspace{0.8 cm}
\newpage
\vspace{0.8 cm}

\newpage

\newpage
\noindent
{\large Figure Captions}
\baselineskip 16pt

\vspace{1.5cm}
\noindent
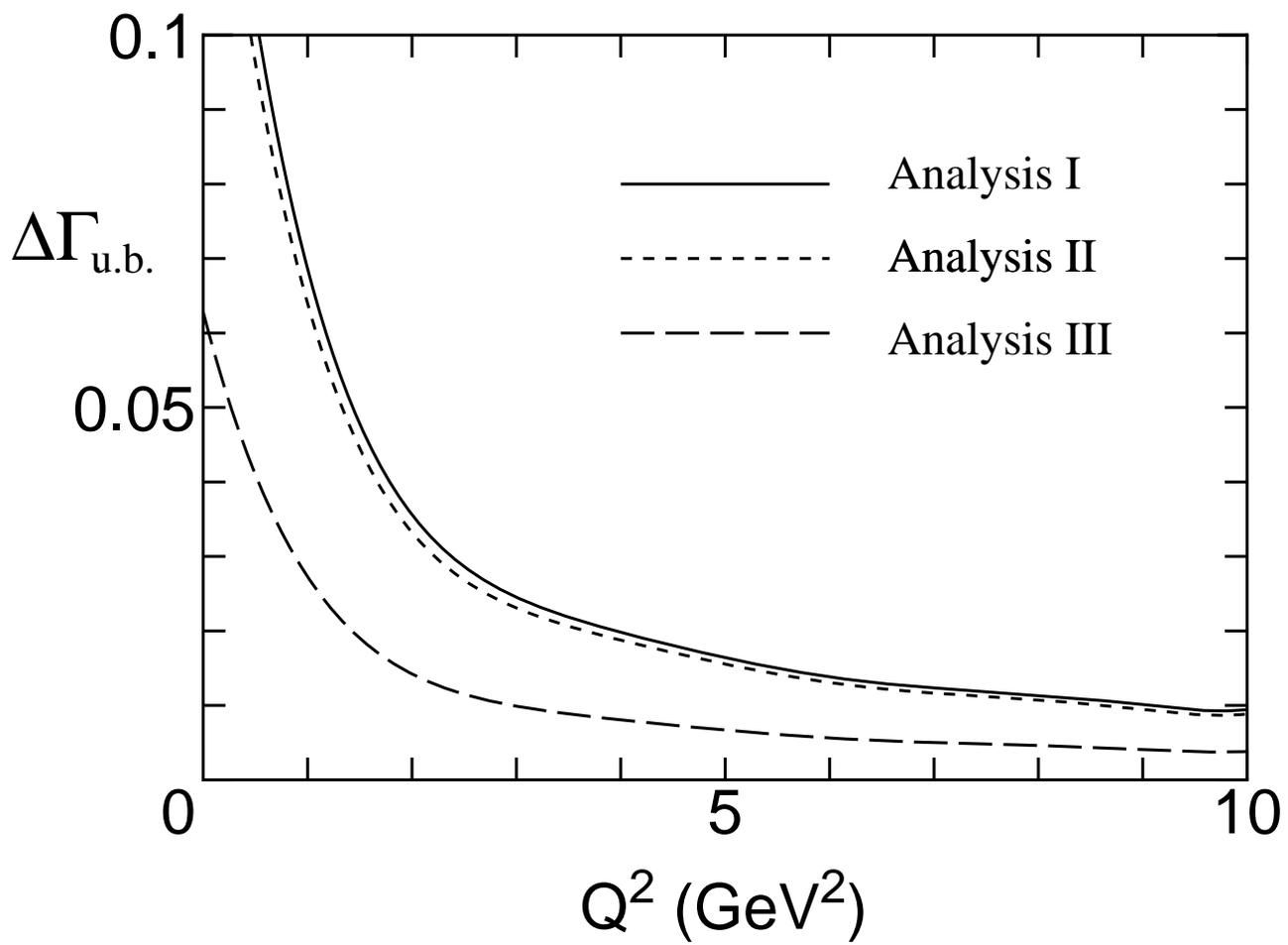
Fig.1 \quad The upper bound for the target mass correction $\Delta\Gamma$,
$\Delta\Gamma_{u.b.}$, as a function of $Q^2$. The solid, short-dashed and
long-dashed lines show the upper bounds for the analyses I, II and III,
respectively.

\vspace{0.5cm}

\noindent
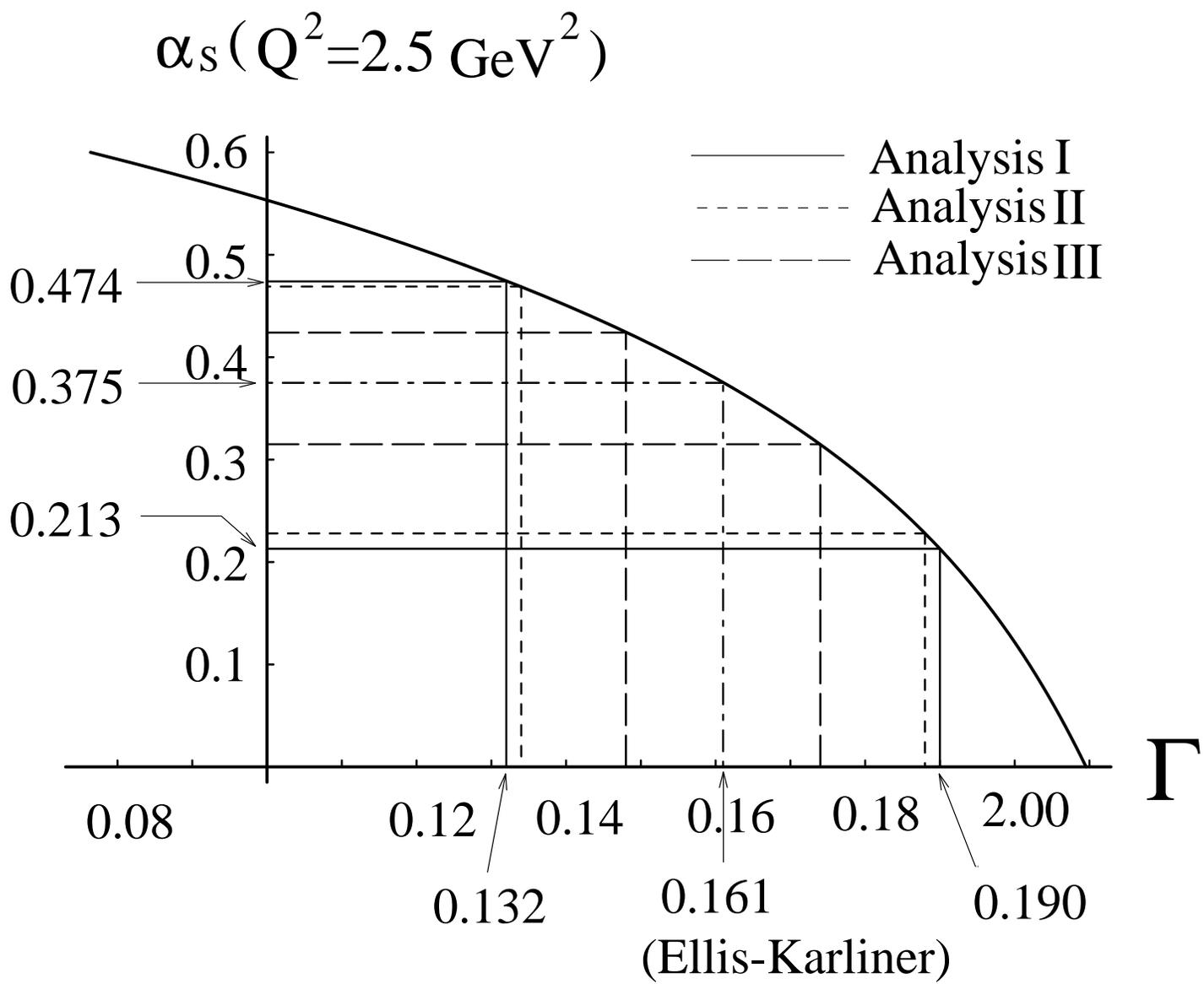
Fig.2 \quad The Bjorken sum rule $\Gamma$ versus the QCD coupling
constant $\alpha_s$ at $Q^2=2.5$GeV$^2$.
The dot-dashed line corresponds to the Ellis-Karliner's analysis.
The solid, short-dashed and long-dashed lines show the upper and lower
limits of $\Gamma$ with the corrections for the analyses I, II and
III, respectively.

\end{document}